\documentclass{emulateapj}
\usepackage{url}

\shorttitle{Constraining the nuclear equation of state of neutron star}
\shortauthors{Lan et al.}
\slugcomment{}

\begin{document}
\title{Constraining the nuclear equation of state via gravitational-wave radiation of short gamma-ray burst
remnants}
\author{Lin Lan\altaffilmark{1,2}, Hou-Jun L\"{u}\altaffilmark{1}, Jared Rice\altaffilmark{3},
and En-Wei Liang\altaffilmark{1}} \altaffiltext{1}{Guangxi Key Laboratory for Relativistic
Astrophysics, School of Physical Science and Technology, Guangxi University, Nanning 530004, China;
lhj@gxu.edu.edu} \altaffiltext{2}{Department of Astronomy, Beijing Normal University, Beijing,
China} \altaffiltext{3}{Department of Physics, Texas State University, San Marcos, TX 78666, USA}

\begin{abstract}
The observed internal plateau of X-ray emission in some short gamma-ray bursts (GRBs) suggests the
formation of a remnant supramassive magnetar following a double neutron star (NS) merger. In this
paper, we assume that the rotational energy is lost mainly via gravitational-wave (GW) radiation
instead of magnetic dipole (MD) radiation, and present further constraints on the NS nuclear
equation of state (EoS) via mass quadrupole deformation and \emph{r}-mode fluid oscillations of the
magnetar. We present two short GRBs with measured redshifts, 101219A and 160821B, whose X-ray light
curves exhibit an internal plateau. This suggests that a supramassive NS may survive as the central
engine. By considering 12 NS EoSs, within the mass quadrupole deformation scenario we find that the
GM1, DD2, and DDME2 models give an $M_{\rm p}$ band falling within the 2$\sigma$ region of the
proto-magnetar mass distribution for $\varepsilon=0.01$. This is consistent with the constraints
from the MD radiation dominated model of rotational energy loss. However, for an \emph{r}-mode
fluid oscillation model with $\alpha=0.1$ the data suggest that the NS EOS is close to the Shen and
APR models, which is obviously different from the MD radiation dominated and mass quadrupole
deformation cases.
\end{abstract}


\section{Introduction}
One favored progenitor model for short gamma-ray bursts (GRBs) is the coalescence of two neutron
stars (NS-NS; Paczynski 1986; Eichler et al. 1989). On 2017 August 17, the first direct detection
of gravitational waves (GWs; GW170817) and {an} electromagnetic counterpart originating from the
merger of a binary NS system was achieved via the collaboration of Advanced LIGO, Advanced VIRGO,
{\em Fermi}, as well as optical telescopes (Abbott et al. 2017a,b; Covino et al. 2017; Goldstein et
al. 2017; Kasen et al. 2017; Savchenko et al. 2017; Zhang et al. 2018). The near-coincident
detection of a short GRB potentially provides the first ``smoking gun" evidence that at least some
short GRBs originate from NS mergers.

The remnant of double NS mergers remains an open question, and is dependent on the total mass of
the post-merger system and the poorly known NS equation of state (EoS; Lasky et al. 2014; Li et al.
2016). One possible remnant of such mergers is a supramassive NS which may survive for seconds to
hours before collapsing into a black hole (BH) if the nascent NS mass is less than the maximum
gravitational mass (Hotokezaka et al. 2013; Zhang 2014; L\"{u} et al. 2015; Foucart et al. 2016;
Gao et al. 2016; Kiuchi et al. 2018). Observationally, a good fraction of the X-ray light curves of
short GRBs were discovered to show an extended plateau, a nearly flat light curve extending to
hundreds of seconds, followed by a sharp decay with a decay index $t^{-(8-9)}$ (called internal
plateau; Rowlinson et al. 2010, 2013; L\"{u} et al. 2015). Such a feature is very difficult to
explain if it is powered by a BH engine, but seems to be consistent with the prediction of a
rapidly spinning, supramassive NS (also called a millisecond magnetar). The sharp decay following
the X-ray plateau is interpreted as the supramassive NS collapsing into a BH after it spins down
due to magnetic dipole (MD) or GW radiation (Usov 1992; Thompson 1994; Dai \& Lu 1998a,b; Zhang \&
M\'esz\'aros 2001; Dai et al. 2006; Gao \& Fan 2006; Metzger et al. 2008; Fan et al. 2013; Zhang
2013, 2014; Ravi \& Lasky 2014; L\"{u} et al. 2015,2017; Gao et al. 2016; Chen et al. 2017).

Previous studies have shown that the newly born supramassive magnetar collapsing into a BH is
triggered by the loss of a large amount of rotational energy due to MD radiation, and have
estimated the physical parameters (i.e., the initial rotation period $P_0$ and the strength of the
dipole magnetic field $B_{\rm p}$) and constrained the NS EoS. (Rowlinson et al. 2010, 2013; Lasky
et al. 2014; L\"{u} et al. 2015). However, they found that the inferred initial rotation period is
much longer than that expected in the double NS-NS merger model (Friedman et al. 1986; Rowlinson et
al. 2013; L\"{u} et al. 2015). Such a puzzle might be solved in two ways. One is a low efficiency
conversion of the magnetar wind energy into radiation, but it seems to be less likely given the
higher expected efficiency of magnetic energy dissipation processes (Drenkhahn \& Spruit 2002; Xiao
\& Dai 2019), or the varying gravitational mass and baryonic mass of the NS (Gao et al. 2019). The
other solution is that most of the rotational energy of the magnetar was carried away via strong GW
radiation (Fan et al. 2013; Lasky et al. 2014). If this is the case, the GW radiation of a newly
born magnetar can be produced via either a mass quadrupole deformation with ellipticity
$\varepsilon$ for an NS rotating as a rigid body or an \emph{r}-mode fluid oscillation with
amplitude $\alpha$ (Owen et al. 1998; Lindblom et al. 1998; Andersson \& Kokkotas 2001; Zhang \&
M\'esz\'aros 2001; Owen 2010; Yu et al. 2010; Fan et al. 2013; Lasky 2015; Ho 2016; Lasky \&
Glampedakis 2016; L\"{u} et al. 2017). These GW signals are too weak to be detected by the current
Advanced LIGO and Advanced Virgo observatories (Alford \& Schwenzer 2014, 2015; Abbott et al.
2017c; L\"{u} et al. 2017, 2019; Ai et al. 2018).

One interesting question is: can a magnetar's rotational energy loss dominated by GW radiation be
used to constrain the NS EoS? In this paper, by analyzing the X-ray emission of short GRBs 101219A
and 160821B, we used the observed data to constrain the EoS of NSs, and then compared the
constraints of a mass quadrupole deformation for different $\varepsilon$ with an \emph{r}-mode
fluid oscillation for different $\alpha$. This paper is organized as follows. The GW radiation
constraints on the NS EoS are presented in Section 2. In Section 3, we show the results of the
constraints for GRBs 101219A and 160821B. The conclusions are drawn in Section 4 with some
discussions. Throughout the paper, a concordance cosmology with parameters $H_0 = 71$ km s$^{-1}$
Mpc $^{-1}$, $\Omega_M=0.30$, and $\Omega_{\Lambda}=0.70$ are adopted.

\section{Constraining the NS EoS via GW radiation dominated rotational energy losses}
The rapid decay after the X-ray plateau in the afterglow of short GRBs indicates that the
supramassive NS is collapsing into a BH. If this is the case, the inferred collapse time can be
used to constrain the NS EoS (Lasky et al. 2014; Ravi \& Lasky 2014; L\"{u} et al. 2015). In this
section, we further constrain the NS EoS by assuming that the loss of rotational energy of the
newly born magnetar is dominated by GW radiation. We will discuss two different scenarios of GW
radiation as follows.

\subsection{GW radiation from mass quadrupole deformation}
The energy reservoir of a millisecond magnetar is the total rotation energy of the NS in rigid
rotation, and it is written as
\begin{eqnarray}
E_{\rm rot} = \frac{1}{2} I \Omega^{2}
\simeq 2 \times 10^{52}~{\rm erg}~
M_{1.4} R_6^2 P_{-3}^{-2},
\label{Erot}
\end{eqnarray}
where $I$, $\Omega$, $P$, $R$, and $M$ are the moment of inertia, angular frequency, rotating
period, radius, and mass of the NS, respectively. The convention $Q = 10^x Q_x$ in cgs units is
adopted. A magnetar loses its rotational energy in two ways: MD torques ($\dot{E}_{\rm em}$) and GW
radiation ($\dot{E}_{\rm gw,q}$),
\begin{eqnarray}
\dot{E}_{\rm rot}= I\Omega \dot{\Omega} &=& \dot{E}_{\rm em} + \dot{E}_{\rm gw, q} \nonumber \\
&=& -{\beta}I\Omega^{4}-{\gamma_{\rm q}}I\Omega^{6},
\label{mountain}
\end{eqnarray}
where $\dot{\Omega}$ is the time derivative of the angular frequency, $\beta= \eta B^2_{\rm
p}R^{6}/6c^{3}I$, and $\gamma_{\rm q}=32GI\varepsilon^{2}/5c^{5}$. $B_{\rm p}$ is the surface
magnetic field at the pole, $\varepsilon=2(I_{\rm xx}-I_{\rm yy})/(I_{\rm xx}+I_{\rm yy})$ is the
ellipticity in terms of the principal moment of inertia, and $\eta$ is the efficiency of converting
the magnetar wind energy into X-ray radiation.

If the magnetar loses most of its rotational energy via GW radiation, one has
\begin{eqnarray}
\dot{E}_{\rm rot}= I\Omega \dot{\Omega} \simeq -{\gamma_{\rm q}}I\Omega^{6},
\label{GW_dominated}
\end{eqnarray}
The full solution of $P(=2\pi/\Omega)$ in Equation (\ref{GW_dominated}) can be written as
\begin{eqnarray}
P(t) &=& P_{0} \left(1+\frac{2048\pi ^{4}}{5}\frac{GI \varepsilon
^{2}}{c^{5}P_{0}^{4}}t\right)^{1/4}\nonumber \\
&=&P_{0} \left(1+\frac{2t}{\tau_{\rm gw,q}}\right)^{1/4}.
\label{Pt:GW}
\end{eqnarray}
where $P_{0}$ is the initial period at $t = 0$ and $\tau_{\rm gw,q}$ is characteristic spin-down
time-scale in this scenario. Following Ho (2016), $\tau_{\rm gw,q}$ can be given as
\begin{eqnarray}
\tau_{\rm gw,q}&=&\left|\frac{E_{\rm rot}}{\dot{E}_{\rm gw,q}}\right|_{\Omega_{0}}
=\frac{1}{2\gamma_{\rm q}\Omega^{4}_{0}}\nonumber \\ &=&1.8\times10^{4}~{\rm
s}~I^{-1}_{45}\varepsilon^{-2}_{-3}P^{4}_{0,-3}.
\label{eq:tgw}
\end{eqnarray}
Based on Equation(\ref{eq:tgw}), one can derive the initial period of the magnetar $P_{0}$,
\begin{eqnarray}
P_{0,-3}=0.1~{\rm s}~I^{1/4}_{45}\varepsilon^{1/2}_{-3}\tau^{1/4}_{\rm gw,q}.
\label{P0:GW}
\end{eqnarray}

For a given EoS, the maximum gravitational mass ($M_{\rm max}$) depends on the period and the
maximum NS mass for a non-rotating NS ($M_{\rm TOV}$). It can be expressed as (Lyford et al. 2003;
Lasky et al. 2014)
\begin{eqnarray}
M_{\rm max} = M_{\rm TOV}\left(1+\hat{\alpha} P^{\hat{\beta}}\right)
\label{Mt1}
\end{eqnarray}
where $\hat{\alpha}$ and $\hat{\beta}$ depend on the NS EoS. The values of $\hat{\alpha}$ and
$\hat{\beta}$ for given EoSs are presented in Table 1.

As the NS spins down, the maximum mass $M_{\rm max}$ gradually decreases. When the proto-magnetar
mass ($M_{p}$) is close to $M_{\rm max}$, the centrifugal force can no longer sustain the
gravitational force and the NS will collapse into a BH. By adopting Equations(\ref{Pt:GW}) and
(\ref{Mt1}), one can derive the collapse time ($t_{\rm col}$) as a function of $M_{\rm p}$ in this
scenario:
\begin{eqnarray}
t_{\rm col} &=& \frac{5}{2048\pi ^{4}}\frac{c^{5}}{GI \varepsilon ^{2}}\left[\left(\frac{M_{\rm p}-
M_{\rm TOV}} {\hat{\alpha} M_{\rm TOV}}\right)^{4/\hat{\beta}}-P_{0}^{4}\right]\nonumber \\
&=&\frac{\tau_{\rm gw,q}}{2P_{\rm 0}^{4}}\left[\left(\frac{M_{\rm p}-M_{\rm TOV}}{\hat{\alpha} M_{\rm
TOV}}\right)^{4/\hat{\beta}}-P_{0}^{4}\right].
\label{tcol:GW}
\end{eqnarray}
For given NS EoS, $M_{\rm TOV}, \hat{\alpha}$, and $\hat{\beta}$ are known. $P_0$ and $t_{\rm col}$
can be inferred from the X-ray observations of short GRBs (more details will be discussed in
section 4). Moreover, the Galactic binary NS population has a tight mass distribution with
$M_p=2.46^{+0.13}_{-0.15}M_\odot$ (Valentim et al. 2011; Kiziltan et al. 2013). Here, we assume
that the distribution of cosmological binary NS masses is the same as that of Galactic binary NS
systems.

\subsection{GW radiation from \emph{r}-mode fluid oscillation}
In the above discussion, it is assumed that the NS undergoes rigid rotation. However, the NS may be
treated as a fluid instead of a rigid body. If this is the case, the dominant GW radiation source
of a newly born magnetar should be the unstable \emph{r}-mode fluid oscillations with amplitude
$\alpha$, whose restoring force is the Coriolis force (Haskell et al. 2015; Lasky 2015). Actually,
GW radiation via the \emph{r}-mode instability of a rotating NS had been discussed in the early
days (Chandrasekhar 1970; Friedman \& Schutz 1978; Andersson 1998; Friedman \& Morsink 1998;
Strohmayer \& Mahmoodifar 2014). The spin-down of a newly born NS can be caused by an r-mode
instability with an oscillating amplitude because of the loss of its angular momentum (Lindblom et
al. 1998; Owen et al. 1998).

Within this scenario, the newly born magnetar spinning down loses its rotational energy via the MD
and \emph{r}-mode GW ($\dot{E}_{\rm gw,r}$) radiation (Owen et al. 1998; Andersson \& Kokkotas
2001; Owen 2010; Ho 2016),
\begin{eqnarray}
\dot{E}_{\rm rot}= I\Omega \dot{\Omega} &=& \dot{E}_{\rm em} + \dot{E}_{\rm gw,r} \nonumber \\
&=& -{\beta}I\Omega^{4}-{\gamma_{\rm r}}I\Omega^{8},
\label{fluid}
\end{eqnarray}
where $\gamma_{\rm r}=(96\pi/15^2)(4/3)^6(GMR^4\tilde{J}^2/c^7\tilde{I})\alpha^2$,
$\tilde{I}=I/MR^2$ and $\tilde{J}$ are dimensionless parameters. Following Alford \& Schwenzer
(2014) and Ho (2016), we adopt the constant values of $\tilde{J}=0.0205$ and $\tilde{I}=0.3$ in our
calculations.

If the NS spins down by losing its rotational energy via GW radiation dominated by the
\emph{r}-mode, one has
\begin{eqnarray}
\dot{E}_{\rm rot}= I\Omega \dot{\Omega} \simeq -{\gamma_{\rm r}}I\Omega^{8},
\label{r-mode_dominated}
\end{eqnarray}
and the full solution of $P(t)$ in Equation(\ref{r-mode_dominated}) can be written as
\begin{eqnarray}
P(t) &=& P_{0} \left[1+\frac{\pi ^{7}}{25}\left(\frac{16}{3}\right)^{6}
\frac{GMR^{4}\tilde{J}^2\alpha^{2}}{c^7\tilde{I}P_{0}^{6}}t\right]^{1/6}\nonumber \\
&=&P_{0} \left(1+\frac{3t}{\tau_{\rm gw,r}}\right)^{1/6}.
\label{Pt:r-mode}
\end{eqnarray}
where  $\tau_{\rm gw,r}$ is the characteristic spin-down time-scale in the \emph{r}-mode scenario;
$\tau_{\rm gw,r}$ can be given by (Ho 2016)
\begin{eqnarray}
\tau_{\rm gw,r}&=&\left|\frac{E_{\rm rot}}{\dot{E}_{\rm gw,r}}\right|_{\Omega_{0}}
=\frac{1}{2\gamma_{\rm r}\Omega^{6}_{0}}\nonumber \\&=&4.3\times10^{5}~{\rm s}~\alpha^{-2}_{-2}P^{6}_{0,-3}.
\label{eq:tgw,r}
\end{eqnarray}
Based on Equation (\ref{eq:tgw,r}), one can derive the rotation period of $P_{0}$ of the magnetar
as
\begin{eqnarray}
P_{0,-3}=0.12~{\rm s}~\alpha^{1/3}_{-2}\tau^{1/6}_{\rm gw,r}.
\label{P0:r-mode}
\end{eqnarray}

By using Equations (\ref{Mt1}) and (\ref{Pt:r-mode}), one can derive the collapse time $t_{\rm
col}$ as a function of $M_{\rm p}$ in this scenario
\begin{eqnarray}
t_{\rm col} &=& \frac{25}{\pi ^{7}}(\frac{3}{16})^{6}\frac{c^{7}\tilde{I}}{GMR^{4}\tilde{J}^2\alpha^{2}}
\left[\left(\frac{M_{\rm p}-M_{\rm TOV}} {\hat{\alpha} M_{\rm
TOV}}\right)^{6/\hat{\beta}}-P_{0}^{6}\right]\nonumber \\
&=&\frac{\tau_{\rm gw,r}}{3P_{\rm 0}^{6}}\left[\left(\frac{M_{\rm p}-M_{\rm TOV}}{\hat{\alpha} M_{\rm
TOV}}\right)^{6/\hat{\beta}}-P_{0}^{6}\right].
\label{tcol:r-mode}
\end{eqnarray}
Similar to Equation (\ref{tcol:GW}), $M_{\rm TOV}$, $\hat{\alpha}$, and $\hat{\beta}$ are known for
a given NS EoS, and $P_{0}$ and $\tau_{\rm gw,r}$ can be inferred from the observations.

\section{Constraining the NS EoS from the observations of short GRBs 101219A and 160821B}
The observed internal plateau of X-ray emission in short GRBs suggests that the central engine of
at least some short GRBs are supramassive NSs (Rowlinson et al. 2010). Here we selected two short
GRBs, 101219A and 160821B, whose X-ray emissions exhibit an internal plateau feature and which have
a measured redshift (Fong et al. 2013; Rowlinson et al. 2013; L\"{u} et al. 2017). Our purpose is
to further constrain the NS EoS by considering GW radiation dominated (mass quadrupole deformation
and \emph{r}-mode fluid oscillation) magnetar energy loss via these two short GRBs.

\subsection{Observations and light-curve fits}
\begin{figure}
\centering
\includegraphics    [angle=0,scale=0.9]     {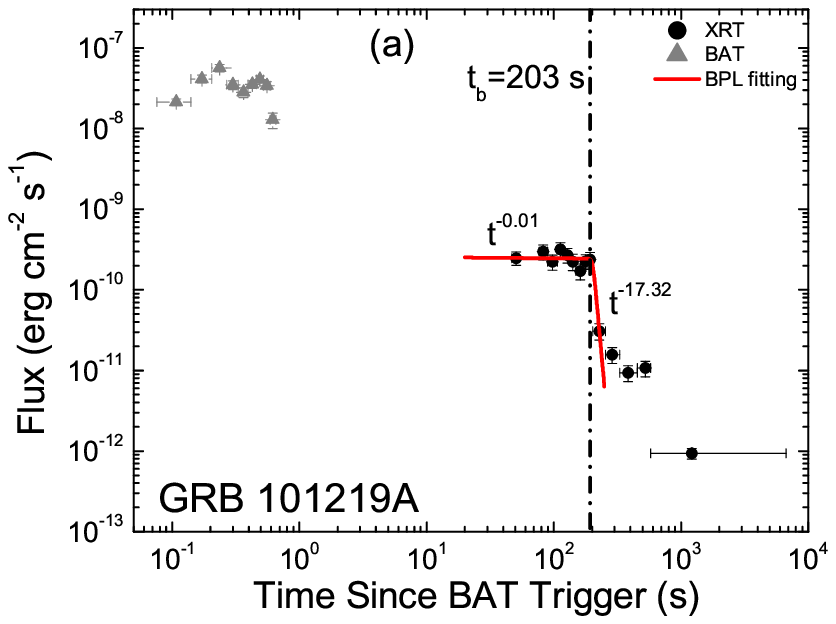}
\includegraphics    [angle=0,scale=0.9]     {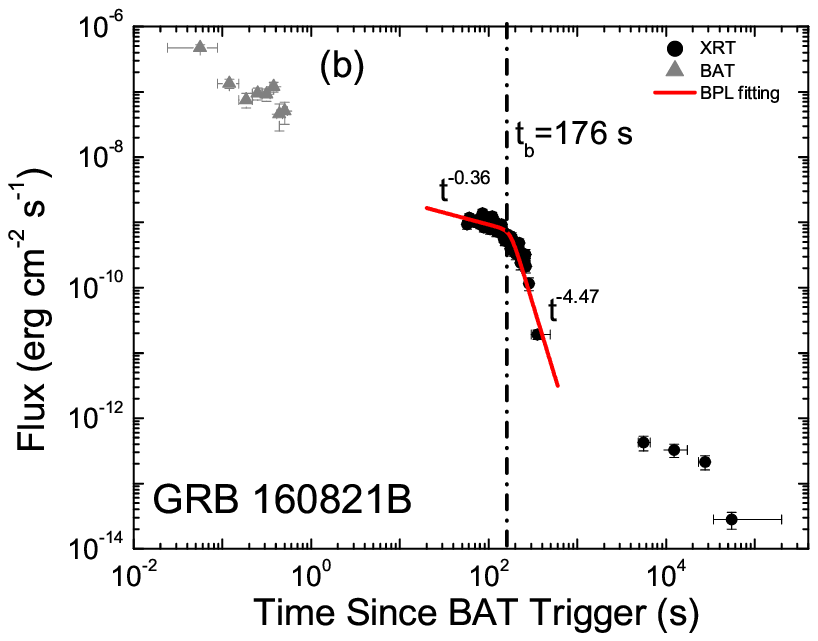}
\caption{X-ray light curves of GRB 101219A (a) and GRB 160821B (b). The red
solid lines show the broken power-law fits, and the black dashed-dotted lines
marked the break time of fits.}
\label{fig:LC}
\end{figure}

The GRB 101219A, triggered by {\em Swift}/Burst Alert Telescope (BAT), is defined as a short GRB
with $T_{90}(15-350~\rm keV)=0.6~\pm~0.2~\rm s$ (Gelbord et al. 2010; Krimm et al. 2010), and a
redshift of $z=0.718$ (Chornock \& Berger 2011). The {\em Swift}/X-Ray Telescope (XRT) began
observation of the GRB field 61 s after the BAT trigger (Golenetskii et al. 2010). The X-ray
afterglow of this GRB presents a plateau emission, followed by a steep decay. More details of the
X-ray light curve are given in Evans et al. (2007, 2009). A smooth broken power-law function is
adopted to fit the X-ray light curve
\begin{eqnarray}
F = F_{0} \left[\left(\frac{t}{t_b}\right)^{\omega\alpha_1}+
\left(\frac{t}{t_b}\right)^{\omega\alpha_2}\right]^{-1/\omega},
\label{BPL}
\end{eqnarray}
where $\omega$ describes the sharpness of the break and is taken to be 3 in this analysis (Liang et
al. 2007). One has $\alpha_{1}=-0.01~\pm~0.09$, $\alpha_{2}=-17.32~\pm~16.79$, and the break time
$t_{b}=203~\pm~22~{\rm s}$ (see Figure \ref{fig:LC}).

GRB 160821B is a nearby short GRB with a redshift of $z=0.16$ (Levan et al. 2016), and was
triggered by {\em Swift}/BAT with $T_{90}(15-350~\rm keV) = 0.48~\pm~0.07~\rm s$ (Palmer et al.
2016; L\"{u} et al. 2017). The XRT began observation of the GRB field 57 s after the BAT trigger
(Siegel et al. 2016), and its X-ray light curve is also characterized by a nearly flat
($\alpha_{1}=-0.36~\pm~0.05$) plateau extending to $t_{b}=176~\pm~3~{\rm s}$, followed by a rapid
decay with $\alpha_{2}=-4.47~\pm~0.23$ (Figure \ref{fig:LC}).

\subsection{Constraining the EoS}
If the rotational energy is lost mainly via GW radiation, one has roughly $t_{\rm
col}=t_{b}/(1+z)$, where $t_{\rm col}$ is the smaller value between $\tau_{\rm gw,q}$ and
$\tau_{\rm gw,r}$. One can then derive the lower limit of $P_{0}$ in the mass quadrupole
deformation and \emph{r}-mode fluid oscillation channels. The only remaining variables in Equations
(\ref{tcol:GW}) and (\ref{tcol:r-mode}) are related to the NS EoS. Here, we consider 12 EoSs that
are usually discussed in the literatures (see Table 1), and EoS parameters are taken from Lasky et
al. (2014), Ravi \& Lasky (2014), Li et al. (2016), and Ai et al. (2018).

\begin{figure}
\centering
\includegraphics    [angle=0,scale=0.45]     {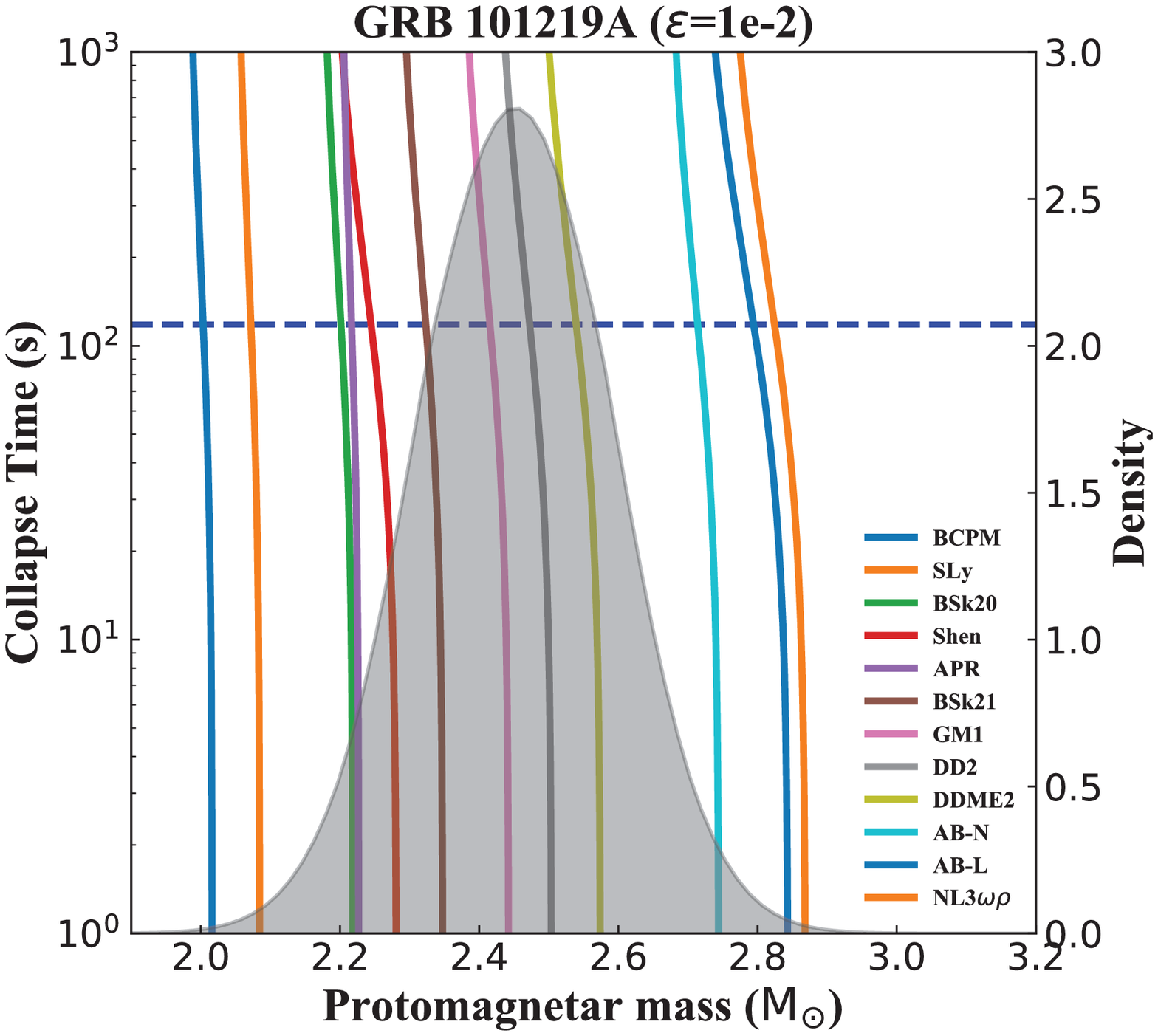}
\includegraphics    [angle=0,scale=0.45]     {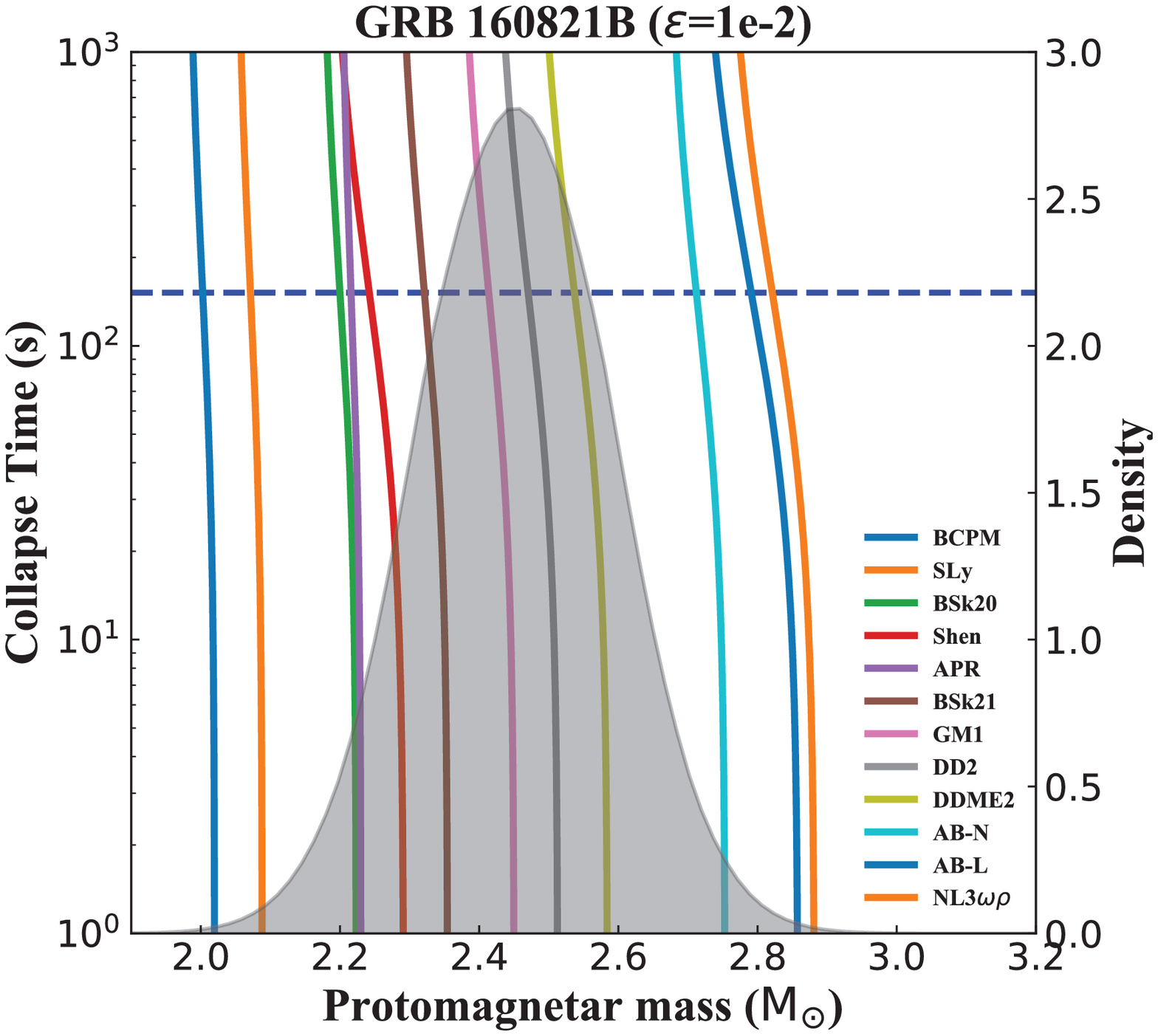}
\caption{Collapse time as a function of the proto-magnetar mass with $\varepsilon=0.01$ for GRBs
101219A and 160821B. The shaded region is the $M_{\rm p}=2.46^{+0.13}_{-0.15} M_{\odot}$
proto-magnetar mass distribution independently derived from the binary NS mass distribution
in the Galactic NS population (Kiziltan et al. 2013). Different colored lines indicate
different EoSs,
and the horizontal blue dotted line is the collapse time.}
\label{fig:mass}
\end{figure}

\begin{figure}
\centering
\includegraphics    [angle=0,scale=0.45]     {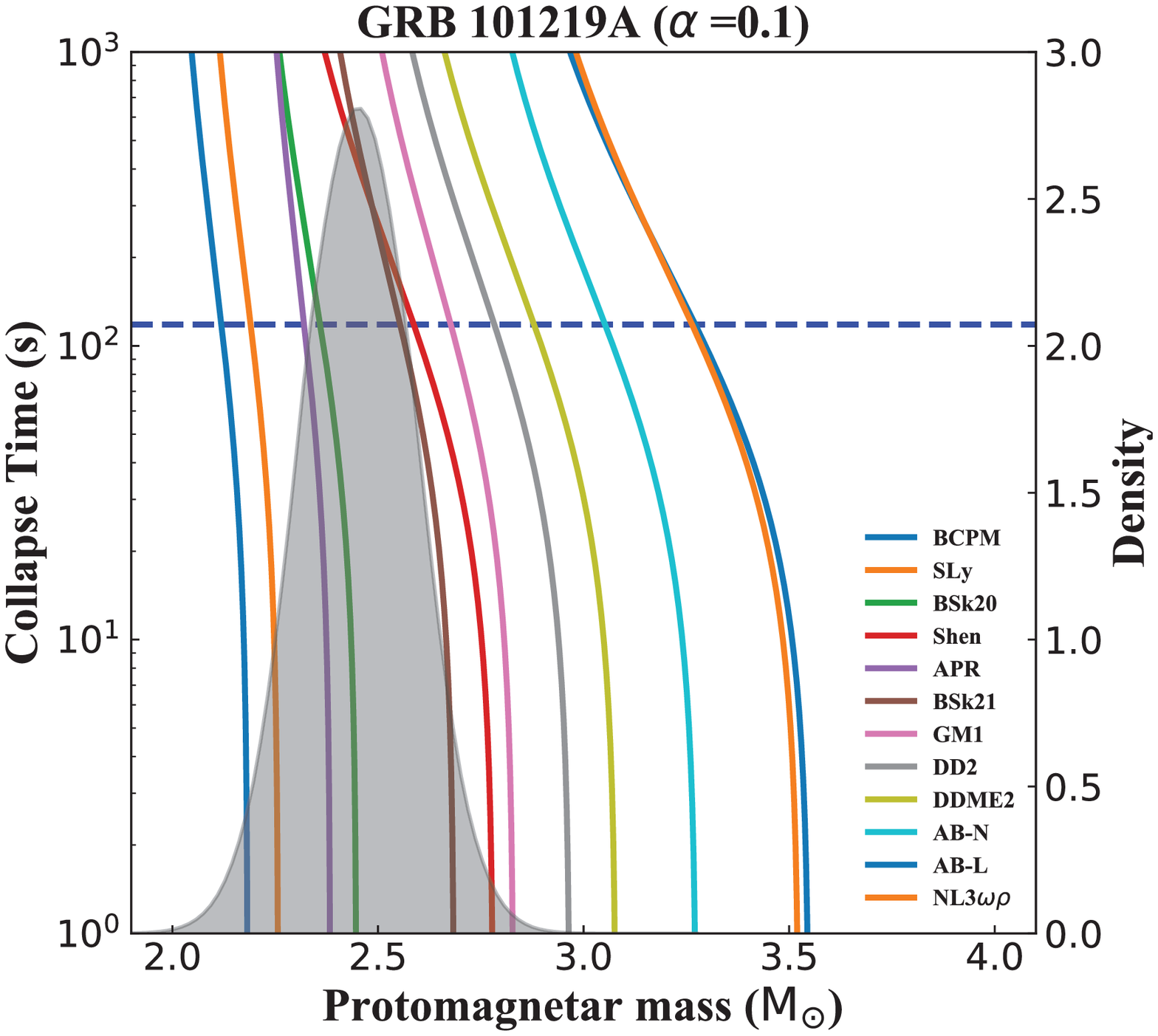}
\includegraphics    [angle=0,scale=0.45]     {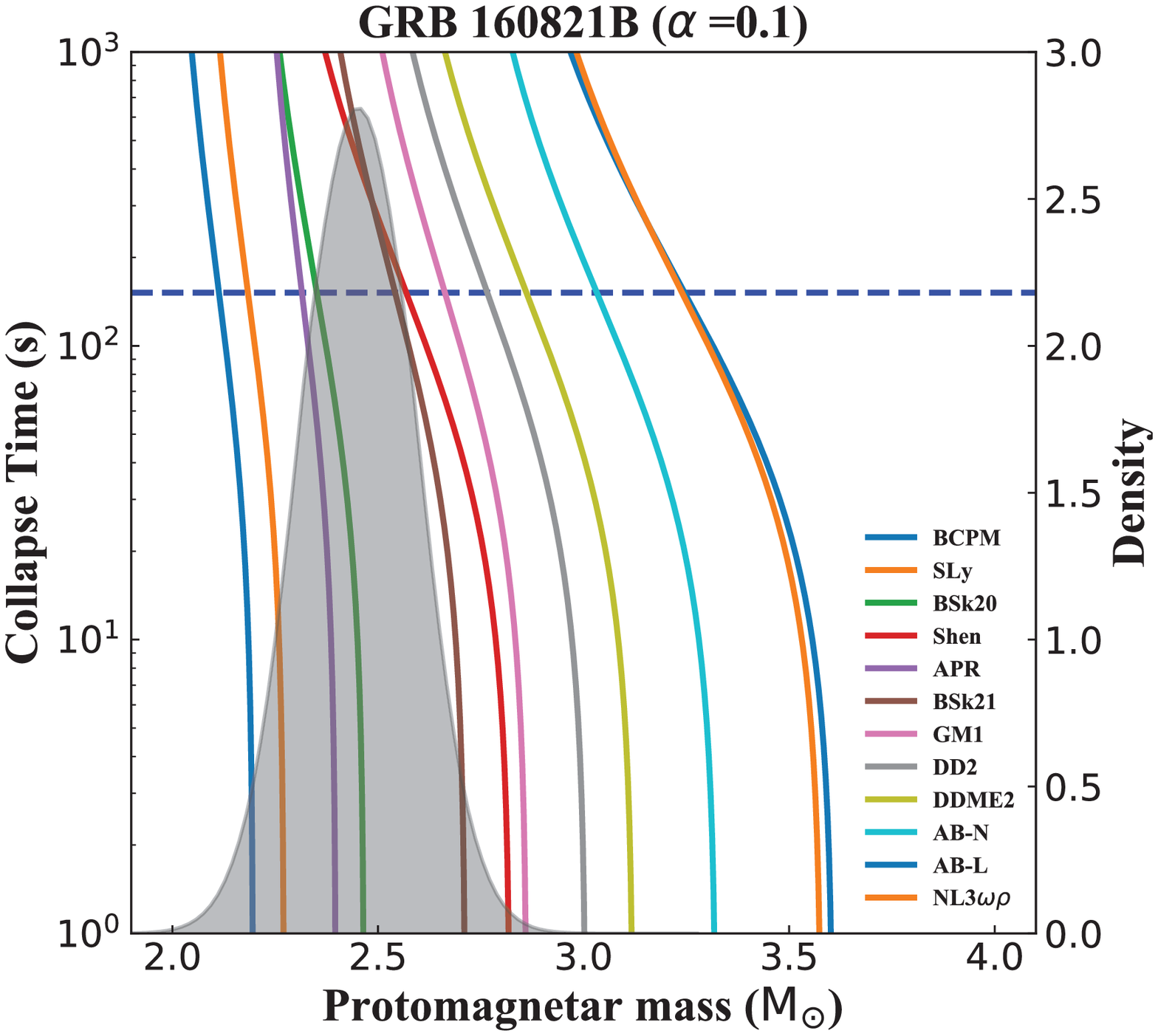}
\caption{Similar to Figure \ref{fig:mass}, but with the \emph{r}-mode fluid oscillation model with
$\alpha=0.1$.}
\label{fig:r-mode}
\end{figure}

Figure \ref{fig:mass} and \ref{fig:r-mode} show the $t_{\rm col}$ as a function of proto-magnetar
mass ($M_{\rm p}$) for GRB 101219A and GRB 160821B, respectively. Different colored lines
correspond to different EoSs. The gray shaded region is the proto-magnetar mass distribution that
is derived independently from the binary NS mass distribution in our Milky Way (Kiziltan et al.
2013; Lasky et al. 2014), and the horizontal dashed lines are the observed collapse time for the
GRBs 101219A and 160821B.

Within the scenario where the GWs are produced by mass quadrupole deformation, the ellipticity
$\varepsilon$ is required to be as large as 0.01 if the rotational energy is lost mainly via GW
radiation (Fan et al. 2013; Lasky et al. 2014; Ho 2016). As such, we adopt $\varepsilon=0.01$ in
our calculations. We find that the GM1, DD2, and DDME2 models give an $M_{\rm p}$ band falling
within the 2$\sigma$ region of the proto-magnetar mass distribution in both GRBs 101219A and
160821B (see Figure \ref{fig:mass}). The correct EoS should be close to those three models, wherein
the maximum mass for a non-rotating magnetar is $M_{\rm TOV}=2.37M_{\odot}$, $2.42M_{\odot}$, and
$2.48M_{\odot}$, respectively. Alternatively, if the GWs are produced by the \emph{r}-mode fluid
oscillation and dominate the energy loss, the \emph{r}-mode amplitude $\alpha$ is required to be as
large as 0.1 (Ho 2016). By adopting $\alpha=0.1$ in our calculations, we find that the Shen and APR
models give an $M_{\rm p}$ band falling within the 2$\sigma$ region of the proto-magnetar mass
distribution in both GRBs 101219A and 160821B (see Figure \ref{fig:r-mode}). The maximum masses for
a non-rotating magnetar in these two models are $M_{\rm TOV}=2.18M_{\odot}$ and $2.20M_{\odot}$,
respectively.

\begin{figure}
\centering
\includegraphics    [angle=0,scale=0.45]     {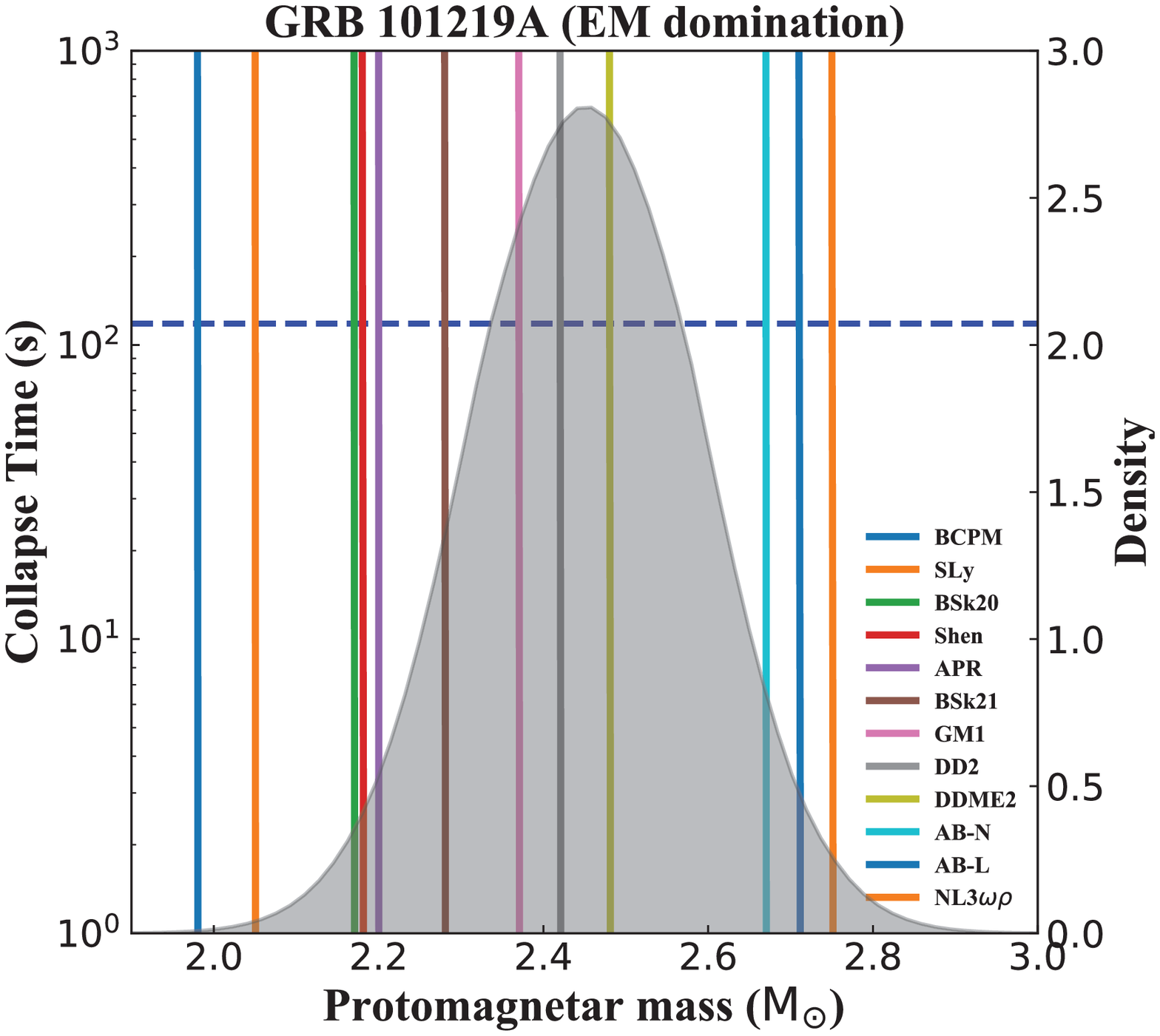}
\includegraphics    [angle=0,scale=0.45]     {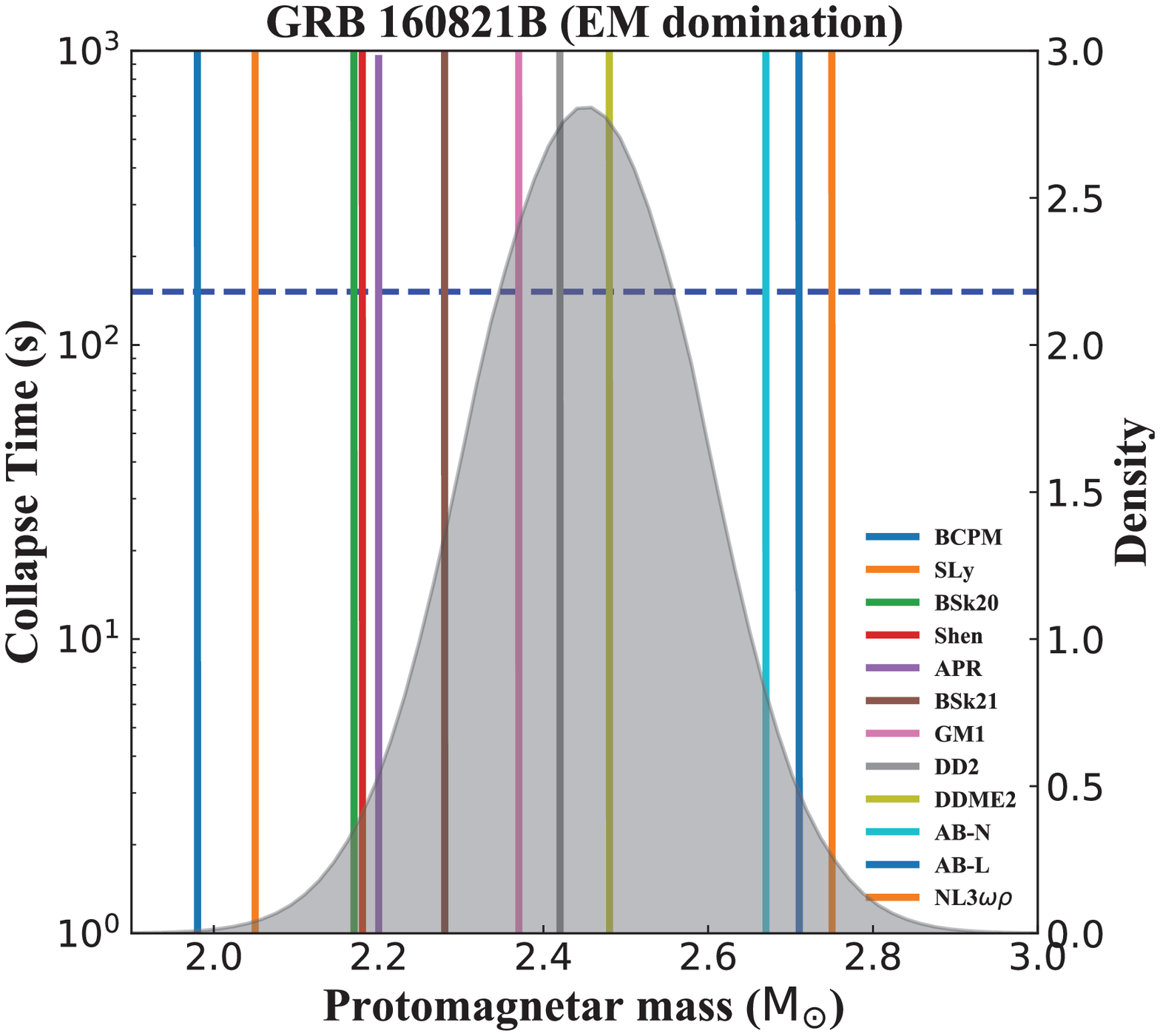}
\caption{Collapse time as a function of the proto-magnetar mass for the MD dominated scenario.}
\label{fig:EM}
\end{figure}

In previous works, within the scenario where the rotational energy of the magnetar is lost mainly
via MD radiation, five NS EoSs are compared to observations to constrain the EoS (Lasky et al.
2014; L\"{u} et al. 2015). In this work in order to compare with the constraints from GW dominated
model, we follow the method of Lasky et al. (2014) and L\"{u} et al. (2015) and 12 EoSs are
compared within the MD radiation dominated model. Figure \ref{fig:EM} presents the collapse time
$t_{\rm col}$ as a function of proto-magnetar mass ($M_{\rm p}$) for GRB 101219A and GRB 160821B,
respectively. We find that the GM1, DD2, and DDME2 models are consistent with the current data.

Comparing the EoS constraints in the MD radiation dominated model with that of the GW dominated
model, we find that the constraints are consistent with each other if the GWs are produced by the
mass quadrupole deformation (Lasky et al. 2014; L\"{u} et al. 2015). However, for the \emph{r}-mode
fluid oscillation model, the constraints are obviously different from that of the MD radiation
dominated model. These results suggest that the NS EoS can be constrained via GW radiation loss of
the the rotational energy of the newly born magnetar, but the constraints are dependent on the
modes of GW radiation (i.e., rigid body rotation or fluid oscillation).

\section{Conclusions and Discussion}
The observed X-ray emission internal plateau in some short GRBs suggests that a supramassive NS may
survive as a remnant of double NS mergers. The supramassive NS then collapses into a BH after a
spin-down via loss of its rotation energy. In this paper, we assume that the rotational energy of
the magnetar is lost mainly via GW radiation, and consider two different GW modes (i.e., mass
quadrupole deformation and \emph{r}-mode fluid oscillation) to constrain the NS EoS via short GRBs
101219A and 160821B. The following interesting results are obtained.
\begin{itemize}
\item We derive the collapse time $t_{\rm col}$ as a function of proto-magnetar mass $M_{\rm
    p}$ by considering the GW radiation dominated energy loss of the NS in the mass quadrupole
    deformation and \emph{r}-mode fluid oscillation modes.
\item The NS EoS can be constrained when the energy loss of the NS is dominated by GW
    radiation, but with the requirement of large $\varepsilon$ (mass quadrupole deformation)
    and $\alpha$ (\emph{r}-mode fluid oscillation) values.
\item Within the scenario where the GWs are produced by the mass quadrupole deformation, the
    constraints on the EoS for the GW radiation dominated model are consistent with that of the
    MD radiation dominated model. The data for short GRBs 101219A and 160821B point toward the
    GM1, DD2, and DDME2 EoSs by assuming $\varepsilon=0.01$. However, for the \emph{r}-mode
    fluid oscillation model with $\alpha=0.1$, the data are different from that of the MD
    radiation dominated model and the correct EoS is closer to the Shen and APR models.

\end{itemize}

The assumed values of $\varepsilon=0.01$ and $\alpha=0.1$ are relatively large in magnitude. It
seems difficult to form a magnetar with these values after a double NS merger. A maximum value of
$\varepsilon=0.001$, as well as a wide range of $\alpha\sim 10^{-5}-0.1$, is constrained for NSs or
other exotic stars (Pitkin 2011; Aasi et al. 2015). Note that the NSs examined in those papers are
much older than the newly born magnetars considered in this work. On the other hand, Lasky \&
Glampedakis (2016) invoke observational data from short GRBs to constrain the ellipticity, and an
upper limit of $\varepsilon$ can be reached at 0.01. Moreover, there is great uncertainty in our
understanding of the physics of \emph{r}-mode and mass quadrupole deformation of NSs (Ho et al.
2011; Lasky \& Glampedakis 2016) and this increases the difficulty of probing the true values of
$\varepsilon$ and $\alpha$.

From the theoretical point of view, a better way of constraining $\varepsilon$ and $\alpha$ is
through detection of a GW signal and its electromagnetic counterpart in a newly born magnetar (Owen
2010; Alford \& Schwenzer 2014). However, even with a large ellipticity and amplitude the GW signal
produced by a newborn rapidly rotating magnetar would be difficult to detect for the current
Advanced LIGO and VIRGO detectors unless the source is nearby; otherwise it may be detected by a
more sensitive instrument in the future, i.e. Einstein Telescope (Alford \& Schwenzer 2014, 2015;
L\"{u} et al. 2017).

Moreover, the main hypothesis of this work is that the energy released by the magnetar during the
plateau phase is caused by GW radiation. Until now, there is no direct evidence to show that
magnetar collapse is caused by GW radiation. However, several lines of indirect evidence suggest
that gGW radiation is one possible energy release mechanism for magnetar collapse. One is that the
inferred initial rotation period is much longer than that expected in the double NS merger model if
the energy release leading to the magnetar collapse is caused by MD radiation (Rowlinson et al.
2013; L\"{u} et al. 2015). The other is that it is possible for a newborn magnetar from a double NS
merger has a larger ellipticity. If this is the case, a surface magnetic field of the magnetar is
required to be as large as $10^{15-16}$ G (Rowlinson et al. 2013; L\"{u} et al. 2015), and the
higher magnetic field can result in a larger NS ellipticity (Gao et al. 2017). L\"{u} et al. (2018)
have shown that a possible signature of GW and MD radiation after the plateau was found in the
X-ray light curve of GRB 060807. In order to test the hypothesis from this work, we hope that GW
radiation associated with a magnetar collapse can be detected by Advanced LIGO and VIRGO in the
future. Then, we can confirm whether or not the magnetar collapse after the double NS merger is
caused by GW radiation.

\begin{acknowledgements}
We acknowledge the use of public data from the {\em Swift} data archive and the UK {\em Swift}
Science Data Center. This work is supported by the National Natural Science Foundation of China
(grant Nos.11922301, 11603006, 11851304 and 11533003), the Guangxi Science Foundation (grant Nos.
2017GXNSFFA198008, 2018GXNSFGA281007, and AD17129006). The One-Hundred-Talents Program of Guangxi
colleges, Bagui Young Scholars Program (LHJ), and special funding for Guangxi distinguished
professors (Bagui Yingcai \& Bagui Xuezhe).

\end{acknowledgements}


\clearpage
\begin{center}
\begin{deluxetable}{cccccccccccccc}
\tablewidth{0pt} \tabletypesize{\footnotesize}
\tablecaption{The parameters of various NS EoS models}
\tablenum{1}

\tablehead{ & \colhead{$M_{TOV}(M_{\odot}$)}& \colhead{\emph{R} (km)} & \colhead{$I(10^{45}~{\rm
g~cm^{2}})$} & \colhead{$\hat{\alpha}(10^{-10}~{s^{-\hat{\beta}}})$} & \colhead{$\hat{\beta}$}}

\startdata
BCPM       &1.98  &9.94   &2.86  &3.39  &-2.65  \\
SLy        &2.05  &9.99   &1.91  &1.60  &-2.75  \\
BSk20      &2.17  &10.17  &3.50  &3.39  &-2.68  \\
Shen       &2.18  &12.40  &4.68  &4.69  &-2.74  \\
APR        &2.20  &10.0   &2.13  &0.303 &-2.95  \\
BSk21      &2.28  &11.08  &4.37  &2.81  &-2.75  \\
GM1        &2.37  &12.05  &3.33  &1.58  &-2.84  \\
DD2        &2.42  &11.89  &5.43  &1.37  &-2.88  \\
DDME2      &2.48  &12.09  &5.85  &1.966 &-2.84  \\
AB-N       &2.67  &12.90  &4.30  &0.112 &-3.22  \\
AB-L       &2.71  &13.70  &4.70  &2.92  &-2.82  \\
NL3$\omega\rho$ &2.75  &12.99  &7.89  &1.706 &-2.88  \\
\enddata

\tablerefs{The parameters of neutron star EoS are taken from Lasky et al.(2014), Ravi \&
Lasky(2014), Li et al.(2016), and Ai et al.(2018).}
\end{deluxetable}
\end{center}



\begin{thebibliography}

\bibitem[Aasi et al.(2015)]{2015ApJ...813...39A} Aasi, J., Abbott, B.~P., Abbott, R., et al.\ 2015,
    \apj, 813, 39


\bibitem[Abbott et al.(2017)]{2017ApJ...848L..13A} Abbott, B.~P., Abbott, R., Abbott, T.~D., et
    al.\ 2017, \apjl, 848, L13


\bibitem[Abbott et al.(2017)]{2017PhRvL.119p1101A} Abbott, B.~P., Abbott, R., Abbott, T.~D., et
    al.\ 2017, Physical Review Letters, 119, 161101


\bibitem[Abbott et al.(2017)]{2017ApJ...851L..16A} Abbott, B.~P., Abbott, R., Abbott, T.~D., et
    al.\ 2017, \apjl, 851, L16


\bibitem[Ai et al.(2018)]{2018ApJ...860...57A} Ai, S., Gao, H., Dai, Z.-G., et al.\ 2018, \apj,
    860, 57


\bibitem[Alford \& Schwenzer(2015)]{2015MNRAS.446.3631A} Alford, M.~G., \& Schwenzer, K.\ 2015,
    \mnras, 446, 3631


\bibitem[Alford \& Schwenzer(2014)]{2014ApJ...781...26A} Alford, M.~G., \& Schwenzer, K.\ 2014,
    \apj, 781, 26


\bibitem[Andersson(1998)]{1998ApJ...502..708A} Andersson, N.\ 1998, \apj, 502, 708


\bibitem[Andersson \& Kokkotas(2001)]{2001IJMPD..10..381A} Andersson, N., \& Kokkotas, K.~D.\ 2001,
    International Journal of Modern Physics D, 10, 381


\bibitem[Chandrasekhar(1970)]{1970PhRvL..24..611C} Chandrasekhar, S.\ 1970, Physical Review
    Letters, 24, 611


\bibitem[Chen et al.(2017)]{2017ApJ...849..119C} Chen, W., Xie, W., Lei, W.-H., et al.\ 2017, \apj,
    849, 119


\bibitem[Chornock \& Berger(2011)]{2011GCN.11518....1C} Chornock, R., \& Berger, E.\ 2011, GRB
    Coordinates Network, Circular Service, No.~11518, \#1 (2011), 11518, 1


\bibitem[Covino et al.(2017)]{2017NatAs...1..805C} Covino, S., Wiersema, K., Fan, Y.~Z., et al.\
    2017, Nature Astronomy, 1, 805


\bibitem[Dai \& Lu(1998)]{1998PhRvL..81.4301D} Dai, Z.~G., \& Lu, T.\ 1998a, Physical Review
    Letters, 81, 4301


\bibitem[Dai \& Lu(1998)]{1998A&A...333L..87D} Dai, Z.~G., \& Lu, T.\ 1998b, \aap, 333, L87


\bibitem[Dai et al.(2006)]{2006Sci...311.1127D} Dai, Z.~G., Wang, X.~Y., Wu, X.~F., \& Zhang, B.\
    2006, Science, 311, 1127


\bibitem[Drenkhahn \& Spruit(2002)]{2002A&A...391.1141D} Drenkhahn, G., \& Spruit, H.~C.\ 2002,
    \aap, 391, 1141


\bibitem[Eichler et al.(1989)]{1989Natur.340..126E} Eichler, D., Livio, M., Piran, T., \& Schramm,
    D.~N.\ 1989, \nat, 340, 126


\bibitem[Evans et al.(2009)]{2009MNRAS.397.1177E} Evans, P.~A., Beardmore, A.~P., Page, K.~L., et
    al.\ 2009, \mnras, 397, 1177


\bibitem[Evans et al.(2007)]{2007A&A...469..379E} Evans, P.~A., Beardmore, A.~P., Page, K.~L., et
    al.\ 2007, \aap, 469, 379


\bibitem[Fan et al.(2013)]{2013PhRvD..88f7304F} Fan, Y.-Z., Wu, X.-F., \& Wei, D.-M.\ 2013, \prd,
    88, 067304


\bibitem[Fong et al.(2013)]{2013ApJ...769...56F} Fong, W., Berger, E., Chornock, R., et al.\ 2013,
    \apj, 769, 56


\bibitem[Foucart et al.(2016)]{2016PhRvD..93d4019F} Foucart, F., Haas, R., Duez, M.~D., et al.\
    2016, \prd, 93, 044019


\bibitem[Friedman et al.(1986)]{1986ApJ...304..115F} Friedman, J.~L., Ipser, J.~R., \& Parker, L.\
    1986, \apj, 304, 115


\bibitem[Friedman \& Schutz(1978)]{1978ApJ...222..281F} Friedman, J.~L., \& Schutz, B.~F.\ 1978,
    \apj, 222, 281


\bibitem[Friedman \& Morsink(1998)]{1998ApJ...502..714F} Friedman, J.~L., \& Morsink, S.~M.\ 1998,
    \apj, 502, 714


\bibitem[Gao et al.(2016)]{2016PhRvD..93d4065G} Gao, H., Zhang, B., \& L{\"u}, H.-J.\ 2016, \prd,
    93, 044065

\bibitem[Gao et al.(2019)]{2019arXiv190503784G} Gao, H., Ai, S.-K., Cao, Z.-J., et al.\ 2019, arXiv
    e-prints, arXiv:1905.03784


\bibitem[Gao et al.(2017)]{2017ApJ...844..112G} Gao, H., Cao, Z., \& Zhang, B.\ 2017, \apj, 844,
    112


\bibitem[Gao \& Fan(2006)]{2006ChJAA...6..513G} Gao, W.-H., \& Fan, Y.-Z.\ 2006, ChJAA, 6, 513


\bibitem[Gelbord et al.(2010)]{2010GCN.11461....1G} Gelbord, J.~M., Barthelmy, S.~D., Chester,
    M.~M., et al.\ 2010, GRB Coordinates Network, Circular Service, No.~11461, \#1 (2010), 11461, 1


\bibitem[Goldstein et al.(2017)]{2017ApJ...848L..14G} Goldstein, A., Veres, P., Burns, E., et al.\
    2017, \apjl, 848, L14


\bibitem[Golenetskii et al.(2010)]{2010GCN.11470....1G} Golenetskii, S., Aptekar, R., Frederiks,
    D., et al.\ 2010, GRB Coordinates Network, Circular Service, No.~11470, \#1 (2010), 11470, 1


\bibitem[Haskell et al.(2015)]{2015ASSP...40...85H} Haskell, B., Andersson, N., D'Angelo, C., et
    al.\ 2015, Gravitational Wave Astrophysics, 40, 85


\bibitem[Ho(2016)]{2016MNRAS.463..489H} Ho, W.~C.~G.\ 2016, \mnras, 463, 489


\bibitem[Ho et al.(2011)]{2011PhRvL.107j1101H} Ho, W.~C.~G., Andersson, N., \& Haskell, B.\ 2011,
    Physical Review Letters, 107, 101101


\bibitem[Hotokezaka et al.(2013)]{2013PhRvD..88d4026H} Hotokezaka, K., Kiuchi, K., Kyutoku, K., et
    al.\ 2013, \prd, 88, 044026


\bibitem[Kasen et al.(2017)]{2017Natur.551...80K} Kasen, D., Metzger, B., Barnes, J., Quataert, E.,
    \& Ramirez-Ruiz, E.\ 2017, \nat, 551, 80


\bibitem[Kiuchi et al.(2018)]{2018PhRvD..97l4039K} Kiuchi, K., Kyutoku, K., Sekiguchi, Y., \&
    Shibata, M.\ 2018, \prd, 97, 124039


\bibitem[Kiziltan et al.(2013)]{2013ApJ...778...66K} Kiziltan, B., Kottas, A., De Yoreo, M., \&
    Thorsett, S.~E.\ 2013, \apj, 778, 66


\bibitem[Krimm et al.(2010)]{2010GCN.11467....1K} Krimm, H.~A., Barthelmy, S.~D., Baumgartner,
    W.~H., et al.\ 2010, GRB Coordinates Network, Circular Service, No.~11467, \#1 (2010), 11467, 1


\bibitem[L{\"u} et al.(2019)]{2019arXiv190406664L} L{\"u}, H.-J., Yuan, Y., Lan, L., et al.\ 2019,
    arXiv:1904.06664


\bibitem[L{\"u} et al.(2015)]{2015ApJ...805...89L} L{\"u}, H.-J., Zhang, B., Lei, W.-H., Li, Y., \&
    Lasky, P.~D.\ 2015, \apj, 805, 89


\bibitem[L{\"u} et al.(2017)]{2017ApJ...835..181L} L{\"u}, H.-J., Zhang, H.-M., Zhong, S.-Q., et
    al.\ 2017, \apj, 835, 181

\bibitem[L{\"u} et al.(2018)]{2018MNRAS.480.4402L} L{\"u}, H.-J., Zou, L., Lan, L., et al.\ 2018,
    \mnras, 480, 4402


\bibitem[Lasky(2015)]{2015PASA...32...34L} Lasky, P.~D.\ 2015, PASA, 32, e034


\bibitem[Lasky \& Glampedakis(2016)]{2016MNRAS.458.1660L} Lasky, P.~D., \& Glampedakis, K.\ 2016,
    \mnras, 458, 1660


\bibitem[Lasky et al.(2014)]{2014PhRvD..89d7302L} Lasky, P.~D., Haskell, B., Ravi, V., Howell,
    E.~J., \& Coward, D.~M.\ 2014, \prd, 89, 047302


\bibitem[Levan et al.(2016)]{2016GCN.19846....1L} Levan, A.~J., Wiersema, K., Tanvir, N.~R., et
    al.\ 2016, GRB Coordinates Network, Circular Service, No.~19846, \#1 (2016), 19846, 1


\bibitem[Li et al.(2016)]{2016PhRvD..94h3010L} Li, A., Zhang, B., Zhang, N.-B., et al.\ 2016, \prd,
    94, 083010


\bibitem[Liang et al.(2007)]{2007ApJ...670..565L} Liang, E.-W., Zhang, B.-B., \& Zhang, B.\ 2007,
    \apj, 670, 565


\bibitem[Lindblom et al.(1998)]{1998PhRvL..80.4843L} Lindblom, L., Owen, B.~J., \& Morsink, S.~M.\
    1998, Physical Review Letters, 80, 4843


\bibitem[Lyford et al.(2003)]{2003ApJ...583..410L} Lyford, N.~D., Baumgarte, T.~W., \& Shapiro,
    S.~L.\ 2003, \apj, 583, 410


\bibitem[Metzger et al.(2008)]{2008MNRAS.385.1455M} Metzger, B.~D., Quataert, E., \& Thompson,
    T.~A.\ 2008, \mnras, 385, 1455


\bibitem[Owen(2010)]{2010PhRvD..82j4002O} Owen, B.~J.\ 2010, \prd, 82, 104002


\bibitem[Owen et al.(1998)]{1998PhRvD..58h4020O} Owen, B.~J., Lindblom, L., Cutler, C., et al.\
    1998, \prd, 58, 084020


\bibitem[Paczynski(1986)]{1986ApJ...308L..43P} Paczynski, B.\ 1986, \apjl, 308, L43


\bibitem[Palmer et al.(2016)]{2016GCN.19844....1P} Palmer, D.~M., Barthelmy, S.~D., Cummings,
    J.~R., et al.\ 2016, GRB Coordinates Network, Circular Service, No.~19844, \#1 (2016), 19844, 1


\bibitem[Pitkin(2011)]{2011MNRAS.415.1849P} Pitkin, M.\ 2011, \mnras, 415, 1849


\bibitem[Ravi \& Lasky(2014)]{2014MNRAS.441.2433R} Ravi, V., \& Lasky, P.~D.\ 2014, \mnras, 441,
    2433


\bibitem[Rowlinson et al.(2013)]{2013MNRAS.430.1061R} Rowlinson, A., O'Brien, P.~T., Metzger,
    B.~D., Tanvir, N.~R., \& Levan, A.~J.\ 2013, \mnras, 430, 1061


\bibitem[Rowlinson et al.(2010)]{2010MNRAS.409..531R} Rowlinson, A., O'Brien, P.~T., Tanvir, N.~R.,
    et al.\ 2010, \mnras, 409, 531


\bibitem[Savchenko et al.(2017)]{2017ApJ...848L..15S} Savchenko, V., Ferrigno, C., Kuulkers, E., et
    al.\ 2017, \apjl, 848, L15


\bibitem[Siegel et al.(2016)]{2016GCN.19833....1S} Siegel, M.~H., Barthelmy, S.~D., Burrows, D.~N.,
    et al.\ 2016, GRB Coordinates Network, Circular Service, No.~19833, \#1 (2016), 19833, 1


\bibitem[Strohmayer \& Mahmoodifar(2014)]{2014ApJ...784...72S} Strohmayer, T., \& Mahmoodifar, S.\
    2014, \apj, 784, 72


\bibitem[Thompson(1994)]{1994MNRAS.270..480T} Thompson, C.\ 1994, \mnras, 270, 480


\bibitem[Usov(1992)]{1992Natur.357..472U} Usov, V.~V.\ 1992, \nat, 357, 472


\bibitem[Valentim et al.(2011)]{2011MNRAS.414.1427V} Valentim, R., Rangel, E., \& Horvath, J.~E.\
    2011, \mnras, 414, 1427

\bibitem[Xiao, \& Dai(2019)]{2019ApJ...878...62X} Xiao, D., \& Dai, Z.-G.\ 2019, \apj, 878, 62


\bibitem[Yu et al.(2010)]{2010ApJ...715..477Y} Yu, Y.-W., Cheng, K.~S., \& Cao, X.-F.\ 2010, \apj,
    715, 477


\bibitem[Zhang et al.(2018)]{2018NatCo...9..447Z} Zhang, B.-B., Zhang, B., Sun, H., et al.\ 2018,
    Nature Communications, 9, 447


\bibitem[Zhang(2014)]{2014ApJ...780L..21Z} Zhang, B.\ 2014, \apjl, 780, L21


\bibitem[Zhang(2013)]{2013ApJ...763L..22Z} Zhang, B.\ 2013, \apjl, 763, L22


\bibitem[Zhang \& M{\'e}sz{\'a}ros(2001)]{2001ApJ...552L..35Z} Zhang, B., \& M{\'e}sz{\'a}ros, P.\
    2001, \apjl, 552, L35




\end{thebibliography}
\end{document}